\documentclass[preprint,showpacs,preprintnumbers,amsmath,amssymb,nofootinbib]{revtex4}

\usepackage{graphicx}

\begin{document}

\preprint{NSF-ITP/02-01}

\title{Classical Black Hole Production in High-Energy Collisions}

\author{Douglas M. Eardley}
\email{doug@itp.ucsb.edu}
\author{Steven B. Giddings}
\email{giddings@physics.ucsb.edu}
\affiliation{
Department of Physics, University of California\\
Santa Barbara, CA 93106-9530
}%

\date{\today}

\begin{abstract}
We investigate classical formation of a $D$-dimensional 
black hole in a high energy collision of two particles.  The existence of
an apparent horizon is related to the 
solution of an unusual boundary-value
problem for Poisson's equation in flat space.  For sufficiently small
impact parameter, we construct solutions giving such apparent horizons in
$D=4$.  These supply improved estimates of the classical cross-section
for black hole production, and of the mass of the resulting black holes.
We also argue that a horizon can be found in a region of weak curvature,
suggesting that these solutions are valid starting points for a
semiclassical analysis of quantum black hole formation.

\end{abstract}

\pacs{Valid PACS appear here}

\maketitle

\section{Introduction}

\def\Mpl{M_{\rm Planck}}
\def\gtwid{\mathrel{\raise.3ex\hbox{$>$\kern-.75em\lower1ex\hbox{$\sim$}}}}
\def\calo{{\cal O}}
\def\beq{\begin{equation}}
\def\eeq{\end{equation}}
\newcommand{\roughly}[1]{\raise.3ex\hbox{$#1$\kern-.75em
\lower1ex\hbox{$\sim$}}}
\def\tu{{\bar u}}
\def\tv{{\bar v}}
\def\tx{{\bar x}}
\def\tz{{\bar z}}
\def\Tt{{\bar t}}
\def\trho{{\bar \rho}}
\def\endsentence{{\hbox{\null}}}

The proposal that the fundamental Planck mass could be as low as a TeV has
excited new 
interest in the problem of black hole formation in
ultra-relativistic collisions.  TeV-scale gravity scenarios offer a
completely new perspective on the hierarchy problem, and 
arise via either
large extra dimensions\cite{Arkani-Hamed:1998rs,Antoniadis:1998ig} 
or compact dimensions with large warp factor; a model for the latter appears
in \cite{Randall:1999ee} and string solutions in \cite{Giddings:2001yu}.  
It has long been believed that high energy collisions where the center
of mass energy substantially 
exceeds the Planck mass would produce black holes;
this statement can be thought of as a simple extrapolation of Thorne's hoop
conjecture\cite{Thorne:1972ji} and has been more recently discussed in
\cite{Banks:1999gd}.  
Lowering the Planck scale to $\calo(TeV)$ thus 
raises the exciting prospect that
black holes can be produced at future accelerators, perhaps even at the 
LHC\cite{Giddings:2001bu,Dimopoulos:2001hw}.\footnote{For a review, see 
\cite{Giddings:2001ih}.}

Clearly we would like to better understand this process.  One important
problem is to estimate the cross-section for black hole production.  It has
been argued that at high energies black hole production has a good
semiclassical description\cite{Banks:1999gd,Giddings:2001bu}, since in such
cases a horizon should form in a region where the curvature is weak and
quantum gravity effects are small.  This leads to the na\"\i ve estimate
that the cross-section for black hole production is roughly given by
\begin{equation}
\sigma\sim \pi r_h^2(\sqrt{s}) \label{crossest}
\end{equation}
where $r_h$ denotes the horizon radius corresponding to CM energy $\sqrt{s}$.  One
would like to make this estimate more precise to improve experimental
predictions, and in particular to derive a differential cross-section
depending on the mass and spin of the black hole produced.  Furthermore,
validity of the estimate (\ref{crossest}) has been 
challenged\cite{Voloshin:2001vs,Voloshin:2001fe}.  In particular, while 
Penrose\cite{Penrose} and D'Eath and Payne\cite{D'Eath:hb,D'Eath:hd,D'Eath:qu}
(for a comprehensive treatment see \cite{D'Eath:book}) have studied the problem
of a
classical collision with zero impact parameter and shown that a closed
trapped surface forms, \cite{Voloshin:2001fe} argues that such collisions
will not form black holes at non-zero impact parameter.
Finally, collisions of cosmic rays with our
atmosphere have energy reach beyond that of LHC, and reasonable
conjectures about neutrino fluxes suggest the possibility that their black hole
products might be seen with present or future cosmic ray 
observatories\cite{Feng:2001ib,Anchordoqui:2001ei,Emparan:2001kf,
Ringwald:2002vk}\cite{Giddings:2001ih}
or absence of their observation could 
place improved bounds on the fundamental Planck
scale\cite{Anchordoqui:2001cg}.  However, unlike the collider setting, where
production above threshhold 
is copious, conclusive statements here depend sensitively on the
exact factors in (\ref{crossest}).  This calls for improved estimates.

This paper will revisit the classical problem of black hole production in
high energy collisions.  In particular, understanding the case of non-zero
impact parameter is crucial to improving the estimate (\ref{crossest}).  
We will use the methods of Penrose\cite{Penrose} and D'Eath and 
Payne\cite{D'Eath:hb,D'Eath:hd,D'Eath:qu}, where each incoming particle is
modelled 
as a point particle accompanied by a plane-fronted gravitational
shock wave, this wave being the Lorentz-contracted longitudinal
gravitational field of the particle.  At the instant of collision
the two shock waves pass through one another, and nonlinearly
interact by shearing and focusing.  Penrose\cite{Penrose} and D'Eath and 
Payne\cite{D'Eath:hb,D'Eath:hd,D'Eath:qu} studied the case of zero impact
parameter $b$, and by finding a closed trapped surface, derived a rigorous
lower bound of $M>\sqrt{s/2}$ and improved estimate $M\approx .84 
\sqrt{s}$ for the
mass of the resulting black hole.  This paper extends this analysis to
$b\neq0$.  In the next section we review the basic approximations involved,
and generalize the construction of
the incoming waves to arbitrary dimensions.  We then reduce the
problem of finding a closed-trapped surface to a rather unusual boundary
value problem for solutions of Poisson's equation; this problem has 
a close analog in the physics of soap
bubbles.  Using conformal methods which apply only in four dimensions, 
section three gives an explicit solution to this problem in that case and
gives a lower bound 
\begin{equation}
  \sigma_{\rm bh\,production} > 32.5(G^2s/4).
\end{equation}
on the classical black hole cross section,
where $G$ is Newton's constant.  Section
four discusses issues in extending this analysis to higher dimensions, and
in better justifying the semiclassical approximation; we follow with
conclusions.

\section{Trapped surfaces in shock-wave geometries}

In the TeV gravity scenarios in question, we will assume that the particles
of the standard model propagate on a brane, but that gravity propagates in
a higher-dimensional space with either large volume or large
warping.\footnote{For a brief unified review of these scenarios, see 
\cite{Giddings:2001ih}.}  This lowers the fundamental Planck scale $M_p$, 
perhaps
to $\calo(TeV)$, while maintaining the observed value of the
four-dimensional Planck mass, $M_4\sim 10^{19} GeV$.  We expect to be able
to create black holes in scattering processes with center of mass energy
$E>M_p$.  In general such black hole solutions will have complicated
dependence on both the gravitational field of the brane, and on the
geometry of the extra dimensions.  However, there are two useful
approximations\cite{Giddings:2001bu} 
that may be used for a wide class of solutions.  First, the
brane is expected to have tension given by roughly 
the Planck or string scale, and so for black holes substantially heavier
than the Planck scale the brane's field should be a negligible effect.
Secondly, if the geometrical scales of the extra dimensions (radii,
curvature radii, variational scale of the warp factor) are all large as
compared to $1/M_p$, then there is a large regime where the geometry of the
extra dimensions plays no essential role.  Therefore it is often a good
approximation to consider high-energy collisions in D-dimensional flat space.

In the center of mass frame, each of the high-energy particles has energy 
\begin{equation}
	\mu = \sqrt s/2\ .
\end{equation}
We use a coordinate system $(\tu,\tv,\tx^i)$
where retarded and advanced times $(\tu,\tv)$ are $(\Tt-\tz,\Tt+\tz)$ in terms
of Minkowski coordinates, $\tz$ being the direction of motion, and $\tx^i$,
$i=1\ldots D-2$, denotes transverse coordinates.  
The impact parameter is $b$ and the particles are initially
incoming along $(\tx^i) = (\pm b/2,0,\ldots,0)$.

The gravitational solution for each of the incoming particles can be found
by boosting the rest-frame solution.  The gravitational field outside a
particle is well approximated by the Schwarzschild solution.  Since we will
be concerned with long-distance phenomena such as formation of large
horizons, short-range modifications of the solution should not be
relevant.  The D-dimensional Schwarzschild solution with mass $M$ is
\begin{equation}
ds^2 = -\left(1- \frac{16\pi GM}{ (D-2) \Omega_{D-2}}\frac{1}{r^{D-3}}\right)
dt^2 + \left(1- \frac{16\pi GM}{(D-2) \Omega_{D-2}}\frac{1}{r^{D-3}}
\right)^{-1} dr^2 + r^2 d\Omega_{D-2}^2\ ,
\end{equation}
where $d\Omega_{D-2}^2$ and $\Omega_{D-2}$ are the line element and volume
of the unit $D-2$ sphere.  The Aichelburg-Sexl
solution\cite{Aichelburg:1970dh} 
is found by
boosting this, taking the limit of large boost and small mass, with fixed
total energy $\mu$.  The result for a particle moving in the $+z$ direction
is the metric
\begin{equation}
ds^2 = -d\tu d\tv + d\tx^{i2} + \Phi(\tx^i) \delta(\tu) d\tu^2\ .\label{AiSe}
\end{equation}
Here $\Phi$ depends only on the transverse radius ${\trho}=
\sqrt{\tx^i\tx_i}$, 
and takes
the form
\begin{eqnarray}
\Phi&=& -8G\mu\ln(\trho)\ ,\ D=4\ ;\label{phifour}\\
\Phi&=& \frac{16\pi G\mu}{\Omega_{D-3}(D-4)\trho^{D-4}}\ ,\ D>4\ .\label{phidef}
\end{eqnarray}
Note that $\Phi$ satisfies Poisson's equation
\begin{equation}
\nabla^2 \Phi = -16\pi G\mu \delta^{D-2}(\tx^i)\ . \label{Poiss}
\end{equation}
in the transverse dimensions 
(here $\nabla$ is the $D-2$-dimensional flat-space derivative in the $(\bar x^i)$).
 This spacetime solution is manifestly
flat except in the null plane $\tu=0$ of the shock wave.  If we consider an
identical shock wave travelling along $\tv=0$ in the $-z$ direction, by
causality these will not be able to influence each other until the shocks
collide.  This means that we can superpose the two solutions of the form
(\ref{AiSe}) to give the exact geometry outside the future light cone of
the collision of the shocks.

The coordinates $\tu,\tv,\tx^i$ suffer the drawback that geodesics and their
tangent vectors appear discontinuous across the shock.  This can be
remedied by going to a new coordinate system defined by
\begin{eqnarray}
\tu &=& u\\
\tv &=& v+\Phi\theta(u) + \frac{u \theta(u) (\nabla\Phi)^2}{4}\\
\tx^i&=& x^i + \frac{u}{2} \nabla_i \Phi(x)\theta(u)\ 
\end{eqnarray}
(where $\theta$ is the Heaviside step function)
in which both geodesics and their tangents are continuous across the shock
at $u=0$.  In these coordinates, the metric of the combined shock waves
becomes
\def\x{{\bf x}}
\def\T#1{\tilde #1}
\def\u{{\bf \T\x}}
\def\del{{\bf \nabla}}
\def\1#1{\frac{1}{#1}}
\def\H#1{H^{(#1)}}
\def\S{{\cal S}}
\def\C{{\cal C}}
\def\N{{\cal N}}
\def\Tr{{\cal T}}
\begin{eqnarray}
 ds^2 &&= -du\,dv + \left[
	\H1_{ik}\H1_{jk} + \H2_{ik}\H2_{jk} - \delta_{ij}
\right] dx^i dx^j			\\
\noalign{\hbox{where}}
  \H1_{ij} &&= \delta_{ij} + \12\nabla_i\nabla_j \Phi(\x-\x_1)\,u\theta(u)
	\label{eq:defH1}\\
  \H2_{ij} &&= \delta_{ij} + \12\nabla_i\nabla_j \Phi(\x-\x_2)\,v\theta(v)
	\label{eq:defH2}\ 
\end{eqnarray}
with $\Phi$ given by eqns.~(\ref{phifour}) and 
(\ref{phidef}), and with 
\begin{eqnarray}
  \x_1 = (+b/2,0,\ldots,0) &&,\quad \x_2 = (-b/2,0,\ldots,0)\ .		
\end{eqnarray}
Here $\x\equiv(x^i)$ in the transverse flat $D-2$-space.

A marginally trapped surface $\S$ is a closed spacelike $D-2$-surface,
the outer null normals of which have zero convergence\cite{HawkEllis}.  For
the case of D=4 and $b=0$, Penrose\cite{Penrose} 
found such a surface in the union of the two shock
waves.  This consisted of two flat disks with radii $\rho_c$ at $\Tt=
-4G\mu\ln\rho_c$, $\tz=\pm4G\mu\ln \rho_c$.  Matching their normals across
the boundary, which lies in the collision surface $u=v=0$, then determined
$\rho_c=4G\mu=r_h$.  
This construction immediately generalizes to the case $D>4$, $b=0$ where
\begin{equation}
	\rho_c = \left(\frac{8\pi G\mu}{\Omega_{D-3}}
\right)^{1/(D-3)}\ . \label  {rhoc}
\end{equation}

Generalizing to $b\neq0$ and arbitrary dimensions,
we attempt to construct $\S$ in the union of the
incoming null hypersurfaces $v\le0=u$ and $u\le0=v$.  These
hypersurfaces intersect each other in the $D-2$-dimensional surface $u=0=v$,
and $\S$ will intersect this $D-2$-surface in a closed $D-3$-dimensional surface
$\C$, to be determined.
In the first incoming null surface $v\le0=u$, $\S$ will be defined by
\begin{equation}
	v = -\Psi_1(\x)\quad\hbox{
	with $\Psi_1>0$ interior to $\C$, $\Psi_1=0$ on $\C$,}
\end{equation}
and one may straightforwardly check that the
outer null normals will have zero convergence for $v<0$
as long as
\begin{equation}
	\nabla^2(\Psi_1  - \Phi_1) = 0 \quad\hbox{interior to $\C$.}\label{difference}
\end{equation}
Similarly, in the second incoming null surface $u\le0=v$, $\S$ will be defined by
\begin{equation}
	u = -\Psi_2(\x)\quad\hbox{
	with $\Psi_2>0$ interior to $\C$, $\Psi_2=0$ on $\C$,}
\label{Psi2def}
\end{equation}
and the outer null normals will have zero convergence for $u<0$
as long as
\begin{equation}
	\nabla^2(\Psi_2  - \Phi_2) = 0 \quad\hbox{interior to $\C$.}
\label{Psi2eqn}
\end{equation}
Finally, the outer null normal to $\S$ must be continuous at
the intersection $u=0=v$; if not there would be a $\delta$-function
in the convergence.  A necessary condition for continuity is
\begin{equation}
	\del\Psi_1 \cdot \del\Psi_2 = 4 \quad\hbox{on $\C$}\ ;
\end{equation}
since $\Psi_{\alpha}$, $\alpha=1,2$,
vanish on $\C$, $\del_i \Psi_{\alpha}$ is normal to $\C$ and
this condition is also sufficient.

Finding a marginally-trapped surface therefore reduces to a simple
mathematical problem.  Specifically, note that (\ref{Poiss}), (\ref{difference})
imply that $\Psi_{\alpha}$ satisfies Poisson's equation with sources at $\x_\alpha$.
Define the rescaled functions
\begin{equation}
g(\x,\x_\alpha;\C) = \frac{\Omega_{D-3}}{16\pi G\mu} \Psi_\alpha
\end{equation}
satisfying 
\begin{eqnarray}
	\nabla_\x^2 g(\x,\x_\alpha;\C) &&= -\Omega_{D-3}\delta^{D-2}(\x-\x_\alpha),	
\label{GFsource}\\
	g(\x,\x_\alpha;\C) &&= 0 \quad\hbox{for $\x$ on $\C$}	\ .\label{DGFdef}
\end{eqnarray}
These are thus the Dirichlet Green's functions 
for sources at $\x_1,\x_2$ and with boundary $\C$.  The problem of finding the
marginally trapped surface is equivalent to the following:

{\sl Problem C. Given two points $\x_1$ and $\x_2$ in Euclidean
$D-2$-space, and a constant $B>0$\hbox{\null}.  Let $g(\x,\x_\alpha;\C)$ be
the Dirichlet Green's functions satisfying (\ref{GFsource}),(\ref{DGFdef}).
Find a closed $D-3$-surface $\C$
enclosing the points with the following property:
\begin{equation}
	\del_\x g(\x,\x_1;\C) \cdot \del_\x g(\x,\x_2;\C) = B^{2}\label{NNN}
\end{equation}
for all points $\x$ on $\C$.}

As a trivial example, if $\x_1=\x_2$, then the unit $D-3$-sphere about
$\x_1$ is a solution $\C$ to Problem C\hbox{\null} for $B=1$\hbox{\null}.
This reproduces Penrose's trapped surface in $D=4$ with suitable scaling,
and gives its generalization, Eq.(\ref{rhoc}), for $D>4$\endsentence.
Given general $\x_1$ and $\x_2$, does a solution $\C$ always exist? Clearly
not if the points are too distant from each other, because a collision
at large enough impact parameter cannot produce a black hole.
Given general $\x_1$ and $\x_2$, is the solution $\C$ unique?  We shall see
that it is not.  We also remark that solutions for different $B$ are
related by simple scale transformations.

One way to understand Problem C is via another physical problem that
serves as a simple analog.  Consider a ring of wire with shape $\C$ in the
$x,y$ plane in three dimensions, and suppose that this ring is spanned by a
soap film.  If we apply pressure to the soap film, then in the limit of
small displacement, its vertical displacement $z(x,y)$ satisfies the
equation
\begin{equation}
\nabla^2 z(x,y) = - \frac{p(x,y)}{\sigma}\ ,\label{presseq}
\end{equation}
where $\sigma$ is the film's tension.  Generalize to the case of two films,
held apart by pressures in the $+z$ and $-z$ directions.  
If the pressure is exerted
at points $\x_\alpha$, then the solutions to (\ref{presseq}) are 
the above Dirichlet
Green's functions.  If the horizontal positions of the pressure points are
the same, and the ring $\C$ is circular, then the angles $\theta_1,\theta_2$
of the soap films
with the $x,y$ plane are constant around $\C$.  Now separate the pressure
points slightly in the $x$-direction 
-- this will change these angles, and they will be
functions of position along $\C$.  Problem C is that of finding the curve
$\C$ for which 
\begin{equation}
B^{2} =\tan\theta_1 \tan\theta_2
\end{equation}
is constant over $\C$, and equal
to the value for zero $x$-separation of the pressure points.  
This problem can also be generalized
to a higher-dimensional analog.  One can argue for the plausibility of a
solution, at least for small enough $x$-displacement, by noting that
deforming a point on $\C$ towards the center increases the angles and thus
the local value of $B$, and deforming away decreases $B$.  This
suggests that $\C$ can be adjusted point-by-point to make $B$ equal to the
given constant value over the ring.  The next section will solve this
problem explicitly in the special case $D=4$.

\section{Explicit construction for $D=4$}

In $D=4$,
Problem C may be readily solved, at least for sufficiently close
points, by a trial-and-correction construction, in two steps:  First, choose 
trial points $\u_1$ and $\u_2$, and a trial curve $\Tr$; then
construct $g(\u,\u';\Tr)$, and evaluate
\begin{equation}
  f(\u;\Tr) \equiv\del_\u g(\u,\u_1;\Tr) \cdot \del_\u g(\u,\u_2;\Tr),
		\quad\hbox{$\u$ on $\Tr$.}
\end{equation}
Second, correct the trial solution by finding a conformal
transformation $\x=\x(\u)$ that
sends $f(\u;\Tr)$ to $B^{2}$; this will send $\u_1$ and $\u_2$ to some
points $\x_1$ and $\x_2$, and will send $\Tr$ to a curve $\C$ obeying
eq.(\ref{NNN}).  Thus we obtain a solution of Problem C\endsentence.
This works because Poisson's equation in dimension 2 is conformally
invariant, while $f$ transforms like
\begin{equation}
  f(\x;\C) =\left|\frac{\partial\u}{\partial\x}\right|^2 f(\u;\Tr).
\end{equation}

Let us now construct some solutions to Problem C\endsentence.
Given two trial points $\u_1$ and $\u_2$ such that
$\frac{1}{2}\left|\u_1 - \u_2\right| \equiv a < 1$; we may take
\begin{equation}
	\u_1 = (a,0), \quad  \u_2 = (-a,0).
\end{equation}
For the trial curve
$\Tr$ choose the unit circle $\left|\u\right| = 1$.
Then the Green function evaluated for the points $\u_1$ and $\u_2$
is
\begin{eqnarray}
	g(\u,\u_1;\Tr) &&= -\12\ln\left(\frac{(\T{x}-a)^2 + \T{y}^2 }{ (a\T{x}-1)^2 + a^2\T{y}^2 }\right),	\\
	g(\u,\u_2;\Tr) &&= -\12\ln\left(\frac{(\T{x}+a)^2 + \T{y}^2 }{ (a\T{x}+1)^2 + a^2\T{y}^2 }\right),
\end{eqnarray}
and
\begin{equation}
	f(\u;\Tr) = \frac{(1-a^2)^2}{(1-2a\T{x}+a^2)(1+2a\T{x}+a^2)}\ .
\end{equation}
To conformally send $f(\u;\Tr)$ to $B^{2}$, we use complex analytic variables
$z\equiv x+iy$ and $\T{z}\equiv \T{x}+i\T{y}$.  We evidently require 
a transformation $z(\T{z})$, analytic on and within the unit circle $\Tr$
parametrized by $\T{z} = \exp(i\T{\phi})$, with
\begin{eqnarray}
  B^{2}\left|\frac{dz}{d\T{z}}\right|^2 = 
	f(\u;\Tr) &&= \frac{(1-a^2)^2}{(1-2a\T{x}+a^2)(1+2a\T{x}+a^2)}	\\
  &&= \frac{(1-a^2)^2}{(1-a^2\exp(2i\T{\phi}))(1-a^2\exp(-2i\T{\phi}))}.
\end{eqnarray}
By inspection, the required tranformation satisfies
\begin{equation}
	\frac{dz}{d\T{z}} = \frac{1-a^2}{B(1-a^2\T{z}^2)}
\end{equation}
which integrates to
\begin{equation}
	z(\T{z}) =
	\frac{1-a^2}{2Ba}\ln\left(\frac{1+a\T{z}}{1-a\T{z}}\right) \ . \label{zxm}
\end{equation}
The true points $(\x_1,\x_2)$ are then found to lie at
\begin{equation}
	x = \pm\frac{1-a^2}{2Ba}\ln\left(\frac{1+a^2}{1-a^2}\right), y=0.
\end{equation}

Restoring physical dimensions,
\begin{eqnarray}
	\Psi_1	&&= 8G\mu g(\x,\x_1;\C),		\\
	\Psi_2	&&= 8G\mu g(\x,\x_2;\C),		\\
	B	&&= \frac{1}{4G\mu},			\\
  \x_1 &&= \left(\frac{2G\mu(1-a^2)}{a}\ln\left(\frac{1+a^2}{1-a^2}\right),0\right)
			= -\x_2,  \\
	b(a)	&&=
	\frac{4G\mu(1-a^2)}{a}\ln\left(\frac{1+a^2}{1-a^2}\right)\ . \label{bofa}  
\end{eqnarray}
Thus we have constructed a marginally trapped surface $\S$ for any
value of impact parameter $b(a)$ that can be obtained from the above formula
for some $a$, with $0\le a<1$.  The area of $\S$ is found to be
\begin{equation}
 {\rm Area}(\S) = 16\pi (G\mu)^2
	\frac{(1-a^2)^2}{a^2}\ln\left(\frac{1+a^2}{1-a^2}\right)\ .
\end{equation}
Now $\S$ may or may not be an apparent horizon, which is defined as the
{\it outermost} marginally trapped surface.  However, the existence
of $\S$ means either that $\S$ is in fact the apparent horizon, or
an apparent horizon exists in the exterior of $\S$\endsentence.  Because
$\S$ can be shown to be convex, and because the 2-metric is Euclidean,
Area$(\S)$ is a lower bound on the area of the apparent horizon.
Modulo technical issues about cosmic censorship, the existence
of an apparent horizon means that the collision will produce
a black hole (or more than one black hole, though this seems unlikely
in the present setup).  Moreover, by the Area Theorem, the
mass of the final black hole (assumed single) is bounded below,
\begin{equation}
	M_{\rm final\,bh} >
  2\mu\frac{1-a^2}{2a}\sqrt{\ln\left({1+a^2 \over 1-a^2}\right)},
		\label{Mbound}
\end{equation}
and the fraction of total energy $2\mu=\sqrt{s}$ emitted as
gravitational radiation is bounded above,
\begin{equation}
	\frac{E_{\rm grav\,rad}}{2\mu} < 1 - 
  \frac{1-a^2}{2a}\sqrt{\ln\left({1+a^2 \over 1-a^2}\right)},
\end{equation}
In fact, $E_{\rm grav\,rad}$ may be significantly smaller because
the final black hole is expected to be rotating (unless $b=0$), tying
up substantial energy.

The function $b(a)$ (for given $\mu$) reaches a maximum value of
\begin{eqnarray}
	b_{\rm max} &&\approx 3.219G\mu		\\
\noalign{\hbox{at}}
	a_{\rm max} &&\approx 0.6153
\end{eqnarray}
so this is the greatest impact parameter for which this construction
can demonstrate production of a black hole.  The corresponding
lower limit on  the cross-section is
\begin{equation}
  \sigma_{\rm bh\,production} \ge \pi b_{\rm max}^2 \approx
  32.552(G\mu)^2.\label{sestim} 
\end{equation}
Previous estimates for the black hole production cross-section used
(\ref{crossest}), which gives 
$\sigma \approx50.27(G\mu)^2$, where $r_h=2G\sqrt{s}=4G\mu$
is the Schwarzschild radius belonging to the total energy available.
Our lower limit is about 65\% of this estimate.
Another interesting quantity is the mass of the final black hole.  
Equation (\ref{Mbound}) together with 
(\ref{bofa}) gives a lower bound on the mass as a function of impact
parameter.  We find a range from $0.71\sqrt{s}$ for $b=0$ to
$0.45\sqrt{s}$
for $b=b_{\rm max}$.  The perturbative analysis of 
\cite{D'Eath:hb,D'Eath:hd,D'Eath:qu}
raises the former to $M\approx .84\sqrt{s}$; we expect a corresponding
increase in the latter upon further analysis.

At values $a>a_{\rm max}$ (but still $a<1$) our construction produces a second,
smaller marginally trapped surface for the same $b$.  This shows that
solutions to Problem C are usually not unique.

\begin{figure*}
\includegraphics{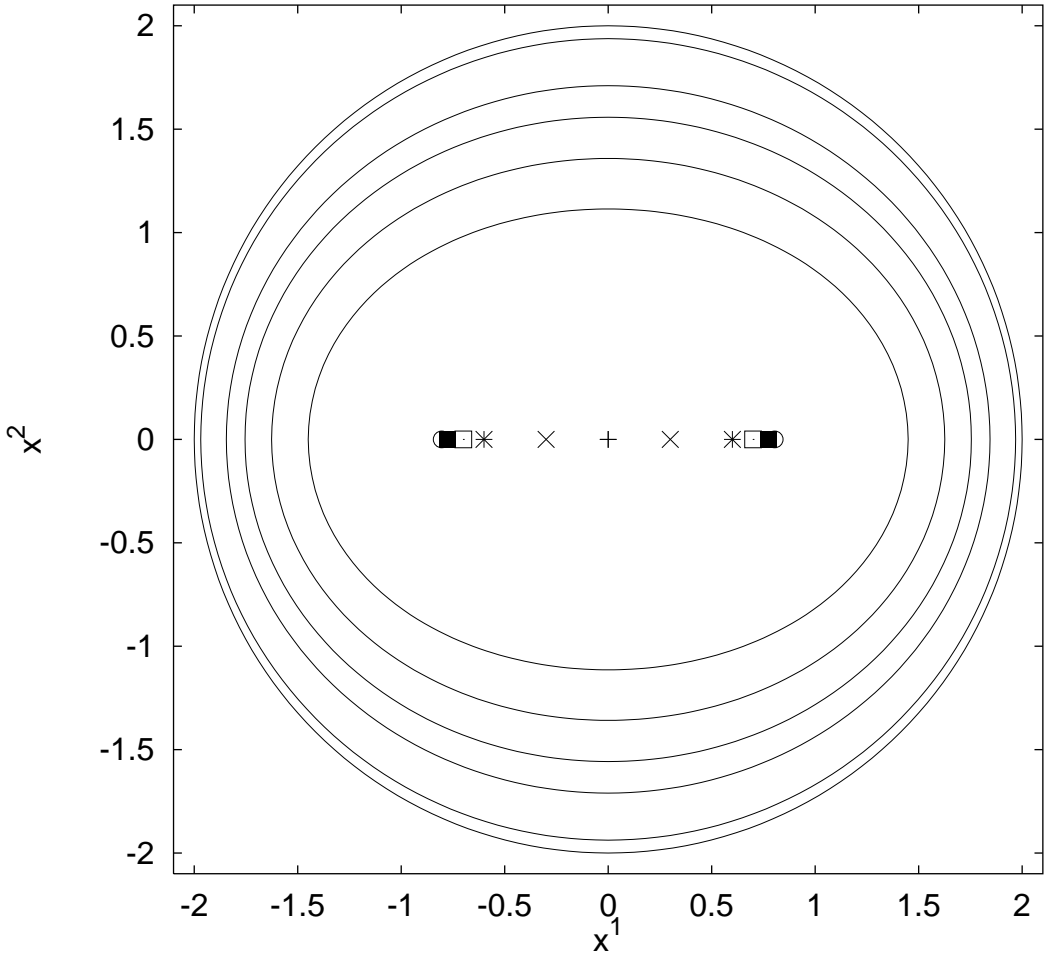}
\caption{\label{apphor}
The intersection curve $\C$ of the marginally trapped surface $\S$ with the
collision plane ($u=0=v$).  Several curves $\C$,
for various impact parameters $b$, are superposed; spacetime dimension
is $D=4$\endsentence.  Distances are in units of $G\sqrt s = 2G\mu$; in these units $r_h=2$.
Incoming particle pairs appear in the horizontal
line $x^2=0$ at pair separation $b$; {\it wider} pairs therefore correspond
to {\it smaller} curves $\C$\endsentence.
Values of $b$ are 0$+$, 0.6$\times$, 1.2$*$, 1.4$\square$, 1.55$\bullet$,
and 1.609$\circ$, the last being maximal.
}
\end{figure*}

It is also interesting to better understand the shape of the curve $\C$ for
the D=4 solution.  This is readily found from (\ref{zxm}), and takes the form
\begin{equation}
(1-a^2) \sinh^2 \frac{ax}{4G\mu(1-a^2)} +
(1+a^2)  \sin^2 \frac{ay}{4G\mu(1-a^2)} = a^2\ .
\end{equation}
Obviously this approaches a circle of radius $4G\mu$ as $a\rightarrow 0$.
Fig.~\ref{apphor} displays the curve $\C$ in the transverse collision plane
for various values of $b$.

\section{Further results and directions}

Clearly it would be desireable to find an explicit solution to Problem C in
the higher-dimensional case, in order to give more careful estimates of the
cross-section in the physically interesting situation where the extra dimensions
are relevant.  Nonetheless, we have given a heuristic argument for the
existence of such a solution, which is buttressed by the explicit
construction of the trapped surface in the four-dimensional case.  This
appears to demonstrate classical black hole formation at non-zero impact
parameter --- answering the criticism of \cite{Voloshin:2001fe}.  

For head-on collisions, $b=0$, in $D>4$, the apparent horizon is the union of two balls of
radius $\rho_c=r_h$, Eq.~(\ref{rhoc}).  The corresponding lower limit on black
hole mass is displayed in Table \ref{tab:mbh}.

\begin{table}
\caption{\label{tab:mbh} Lower limit on $M_{BH}$ in head-on collisions $b=0$,
for various dimensions $D$.}  Here $A_{\rm Sch}$ is the ($D-2$-dimensional)
horizon area of a Schwarzschild black hole of mass $\sqrt s$,
while $A_{\rm trap}$ is the area of the Penrose marginally trapped surface.
\begin{ruledtabular}
\begin{tabular}{rrrr}
 D  & $A_{\rm trap}/A_{\rm Sch}$ &    $M_{BH}/\sqrt s$  &      $1-M_{BH}/\sqrt s$\\
\hline
    4 &      0.50000 &      0.70711 &      0.29289 \\
    5 &      0.54270 &      0.66533 &      0.33467 \\
    6 &      0.55032 &      0.63894 &      0.36106 \\
    7 &      0.55080 &      0.62057 &      0.37943 \\
    8 &      0.54928 &      0.60696 &      0.39304 \\
    9 &      0.54720 &      0.59642 &      0.40358 \\
   10 &      0.54502 &      0.58798 &      0.41202 \\
   11 &      0.54293 &      0.58105 &      0.41895 \\
\end{tabular}
\end{ruledtabular}
\end{table}

Solving Problem C in $D>4$ for $b>0$ may require numerical work.  Note that by
symmetry it still remains a two-dimensional problem of finding a curve
${\hat\C}$ which produces the surface $\C$ as a surface of revolution about
the axis connecting the two source points.  However, in the $D>4$ case, the
relevant Green function is of the form (\ref{phidef}) and no longer transforms
nicely under conformal transformations.  Of course even solving this
problem is only a starting point -- it provides a lower bound on the mass
of the black hole, but as was found in the axial symmetric case in 
\cite{D'Eath:hb,D'Eath:hd,D'Eath:qu}, the resulting black hole will absorb
more of the energy contained in its external fields, and it should be 
possible to raise this
bound by studying this subsequent evolution.  Recall that in the
axial-symmetric case in $D=4$ this resulted in an additional enhancement of
approximately $119\%$.  Going beyond this, the ultimate goal of such an
analysis -- and its quantum extension -- is to compute the differential
cross-section depending on the mass and spin of the resulting black hole.

One might also wonder if a black hole is produced in the high energy collision
of a particle with a purely gravitational shock wave, in any dimension $D\ge4$.
The present analysis suggests that the
answer is ``no".   If we attempt to replace $\Phi_2(\x)$ by a source-free
solution of Laplace's equation, to model a purely gravitational shock wave,
then Eqs.(\ref{Psi2def},\ref{Psi2eqn}) have no solution at all, by
the maximum principle for elliptic equations.  Therefore no apparent
horizon exists in the incoming wavefront surface.  Similarly, the collision
of two purely gravitational shock waves \cite{KhanPenrose} seems not to produce
a black hole.  Moreover, in the collision of two particles at $b>0$,
as studied here, one might wonder if additional, smaller marginally trapped
surfaces might appear, enclosing one particle but not the other; a similar
argument shows not.  We have not considered trapped surfaces that might
exist to the future of the incoming wave surface, however, so these arguments
cannot conclusively rule out black holes.

It is also interesting to compare the estimate (\ref{sestim}) to a heuristic
argument presented in \cite{Anchordoqui:2001cg}.  Anchordoqui et. al. argue
that one may improve estimates of the cross-section by taking into account
angular momentum dependence of the Schwarzschild radius.  Specifically, for
CM energy $\sqrt{s}$ and impact parameter $b$, the angular momentum is
$J=b\sqrt{s}/2$.  One expects that the maximum impact parameter occurs near a
value of $b$ that equals the corresponding angular-momentum dependent
radius $r_h$.  This is given by\cite{Myers:un}
\begin{equation}
r_h^{D-5}\left(r_h^2+\frac{(D-2)^2J^2}{4M^2}
\right) = 
\frac{16 \pi G M}{(D-2) \Omega_{D-2} }
\quad \buildrel J \rightarrow 0 \over\longrightarrow \quad
r_h^{D-3}=\frac{16 \pi G M}{(D-2) \Omega_{D-2} }\ .
\label{rsch}
\end{equation}
If we set $b=r_h$ in the former expression, that gives
\begin{equation}
r_h^{D-3}= \frac{16 \pi G M}{(D-2) \Omega_{D-2} }\left[1+
\frac{(D-2)^2}{16}\right]^{-1}\ .
\end{equation}
This leads to a cross-section estimate
\begin{equation}
\sigma\approx \pi r_h^2 = \left[1+\frac{(D-2)^2}{ 16}\right]^{-\frac{2}{D-3}}
\pi r_{h,{\rm spherical}}^2\ .
\end{equation}
In D=4, this gives a relative correction factor of $64\%$, closely
matching the factor in (\ref{sestim}).  The correspondence of these results suggests
that this estimation technique
 may indeed be approximately correct in higher dimensions.

A final point concerns the validity of the semiclassical approximation.
We expect that the semiclassical approximation for black hole formation
should be justified if a horizon forms at small curvature.\footnote{In the
context of string theory, the Schwarzschild radius must also be larger than
the string length $\sqrt{\alpha'}$\cite{Amati:1987uf,Amati:1988tn}.}
Of course the
surface ${\cal S}$ that we have constructed lies in the plane of the
shocks, and thus is in a sense close to a region of large curvature;
understanding the importance of this would require regulating the solution
taking into account finite size/mass effects.  However, we believe that
this is not necessary, as it is possible to find a trapped surface
outside of the planes of the incoming shocks.
First, in the case $b=0$,
consider Penrose's flat disk $\S$ of radius $\trho=\rho_c$
in the incoming null wavefront surface $\tu=0$.
Construct the null plane $\N$ emanating from
$\S$ in the opposite direction $v=-{\rm const}$; $\N$ has
zero convergence because it is a null plane.  Deform
$\S$ into the future along $\N$, while
leaving it fixed in some neighborhood of its boundary $\C$\endsentence.  This
deformed $D-2$-surface $\S'$ will still have zero convergence along
the $v$-direction everywhere, and thus will be an apparent
horizon, now with weak spacetime curvature on it.  (In fact, $\S'$
has exactly the same area as $\S$, and so gives exactly
the same bound on black-hole mass.)

Second, in the case $b>0$, we can proceed similary.
We can still construct
a null surface $\N$ emanating to the future from our original
marginally trapped surface $\S$, generated by null geodesics
normal to $\S$\endsentence.  However, $\N$ will not be a flat null
plane, because the null normals to $\S$ have nonzero shear.  We
can still deform $\S$ some distance to the future along $\N$
to create a new $D-2$-surface $\S'$ but the convergence
of the null normals of $\S'$ will go positive, due to the shear
(we cannot go too far or we will run into a caustic, i.e. 
the convergence will go to $+\infty$).  Thus $\S'$ will actually
be a trapped surface, not a marginally trapped surface. This
would mean that a marginally trapped surface must lie
somewhere outside of it.  In any case $\S'$ has weak spacetime curvature
everywhere on it, and implies the existence of a black hole.

\section{Conclusions}

The existence of a closed trapped surface in the collsion geometry of two
ultra-relativistic particles clearly demonstrates classical black hole
formation.  The argument that these surfaces are present in the weak
curvature region further suggests that this process can be consistently
treated in a semiclassical analysis, and should help lay the foundation for a
more rigorous justification of such an analysis.  Furthermore, we have
found improved estimates on the production cross-section for black holes.
While these estimates are not enormously different from the more na\"\i ve
estimates of \cite{Giddings:2001bu,Dimopoulos:2001hw}, it is important to
know their size in improving discussion of the sensitivity of cosmic ray
observations to black hole production.  Future directions include more
complete analysis of the higher-dimensional classical problem, which may
require numerical work.  The present classical analysis should serve 
as a
starting point for a more complete investigation of the semiclassical
approximation to black hole formation.

After completing this work we were kindly informed by 
R.~Penrose that he had found related unpublished results on the existence
of a maximal impact parameter
in
$D=4$, many years ago.

\begin{acknowledgments}
Conversations with P.~D'Eath, J.~Hartle, G.~Horowitz, S.~Hughes, L.~Lindblom,
R.~Penrose,
J.~Polchinski, M.~Scheel, A.~Shapere, K.~Thorne, and S.~Zelditch
are gratefully acknowledged.
This research was supported in part by the National Science Foundation 
under Grant No.\ PHY99-07949 (DME) and by the DOE under contract
DE-FG-03-91ER40618 (SBG).  
\end{acknowledgments}

\bibliography{paper}

\end{document}